\documentclass[10pt,leqno]{amsart}
\usepackage{graphicx}
\input pz.sty
\input epsf.sty

\begin{document}
\newcommand{\apjs}{Astrophys.~J. Suppl. }
\setcounter{page}{1}
\PZhead{5}{44}{2024}{10 September}{17 September}
DOI: 10.24412/2221-0474-2024-44-50-54

\bigskip

\PZtitle{Colour-Based Disentangling of Mira Variables and Ultra-Cool Dwarfs}

\PZauth{Aleksandra Avdeeva$^1$, Kefeng Tan$^2$, Santosh Joshi$^3$, Dana Kovaleva$^1$,
Harinder P. Singh$^4$, Ali Luo$^2$, Oleg Malkov$^1$}

\PZinst{Institute of Astronomy of the Russian Academy of Sciences, 48 Pyatnitskaya St., Moscow 119017, Russia}


\PZinst{CAS Key Laboratory of Optical Astronomy, National Astronomical Observatories, Chinese Academy of Sciences, Beijing 100101, China}

\PZinst{Aryabhatta Research Institute of Observational Sciences, Manora Peak, Nainital 263002, India}

\PZinst{Department of Physics \& Astrophysics, University of Delhi, Delhi 110007, India}



\PZbegintext{Avdeeva et al.: Disentangling of Mira Variables and Ultra-Cool Dwarfs}

\small{Despite having different astronomical characteristics, the studies of mira variables and ultra-cool dwarfs frequently show similar red colors, which could cause leading to photometric misclassification. This study uses photometric data from the WISE, 2MASS, and Pan-STARRS surveys to construct color-based selection criteria for red dwarfs, brown dwarfs, and Mira variables. On analyzing the color indices, we developed empirical rules that separate these objects with an overall classification accuracy of approximately 91\%-92\%.
While the differentiation between red dwarfs and both Mira variables and brown dwarfs is effective, challenges remain in distinguishing Mira variables from brown dwarfs due to overlapping color indices. The robustness of our classification technique was validated by a bootstrap analysis, highlighting the significance of color indices in large photometric surveys for stellar classification.}

\bigskip

\section{Introduction}

A Mira-like variable is a type of pulsating variable star, primarily characterized by its long-period variability and significant changes in brightness. These stars, typically late-type red giants on the asymptotic giant branch (AGB), exhibit periodic pulsations due to the expansion and contraction of their outer layers. The periods of these pulsations generally range from 80 to over 1,000 days, with amplitude variations that can reach several magnitudes. Mira-like variables include both Mira variables, which have well-defined periods and large amplitude changes, and semi-regular variables (SRVs), which show less regular periodicity and smaller amplitude variations. These stars are important astrophysical objects for studying the late stages of stellar evolution, as their variability provides insights into the physical processes occurring in evolved stars.

An ultra-cool dwarf is a type of star or brown dwarf with an effective temperature below approximately 2700 K, placing it among the coolest stellar and substellar objects in the universe. These objects include the latest spectral types M, L, T, and Y. Due to their low temperatures, these objects exhibit complex atmospheric chemistry, including the formation of condensate clouds. Ultra-cool dwarfs bridge the gap between the smallest stars capable of sustaining hydrogen fusion and the more massive planets, making them significant in studies of both stellar and planetary formation and evolution.

Mira variables and ultra-cool dwarfs, despite their distinct physical properties, can sometimes be confused in observations due to their low temperatures and resulting red colors, which make both appear as red objects in the night sky. This similarity can complicate their differentiation. Without detailed spectral analysis, which reveals the pulsation characteristics of Mira variables or the distinct molecular absorption features of ultra-cool dwarfs, these objects may be misclassified based on limited photometric data alone. 

These article aims at establishing the colour selection criteria for differentiation between Mira variables and ultra-cool dwarfs based on their colour indices solely. We use the photometric bands of WISE \cite{2012wise.rept....1C}, 2MASS \cite{2003yCat.2246....0C} and Pan-STARRS \cite{2016arXiv161205560C} surveys for this task. 

\section{Data}

We used \cite{2018ApJS..234....1B} as the source for ultra-cool dwarfs, including M red dwarfs and L\&T brown dwarfs. This source provides WISE, 2MASS, and Pan-STARRS magnitudes for 1,601 brown dwarfs and 8,234 red dwarfs. For the Mira variables, we selected a random sample from the Simbad\footnote{https://simbad.u-strasbg.fr/simbad/} database. We cross-matched this sample with the aforementioned surveys and obtained data for slightly more than 11000 Mira variables 

\begin{figure}[ht]
    \centering
    \includegraphics[width=0.48\textwidth]{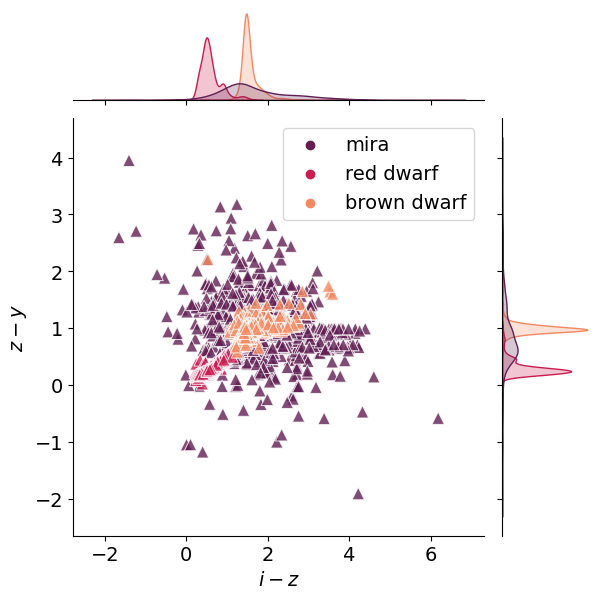} 
    \includegraphics[width=0.48\textwidth]{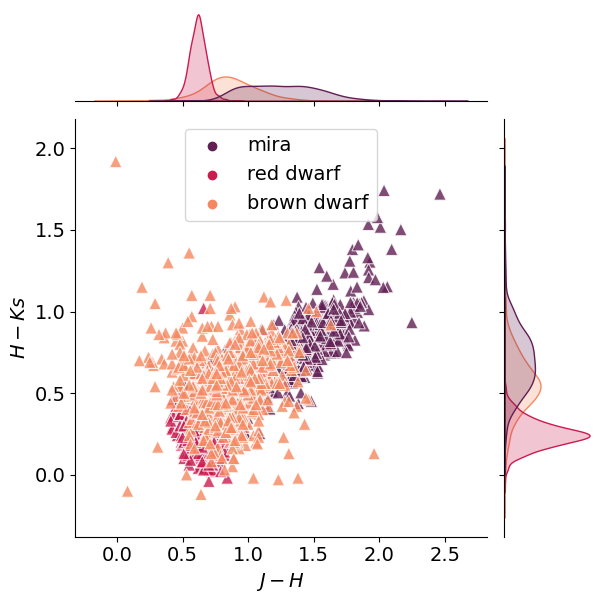}    \\
    \includegraphics[width=0.48\textwidth]{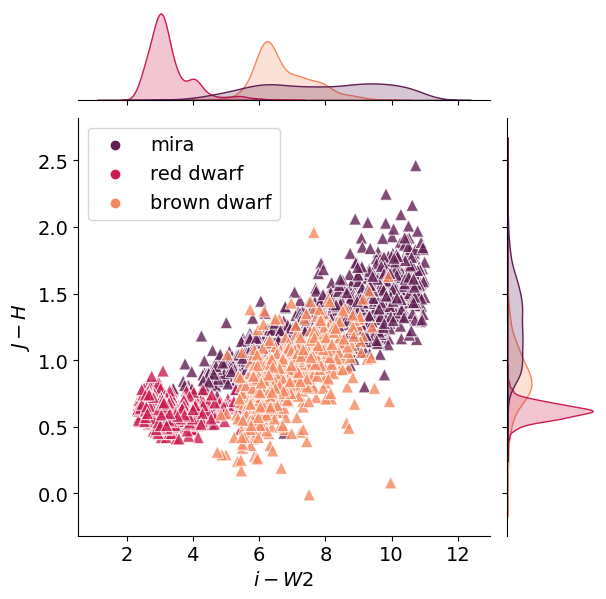} 
    \includegraphics[width=0.48\textwidth]{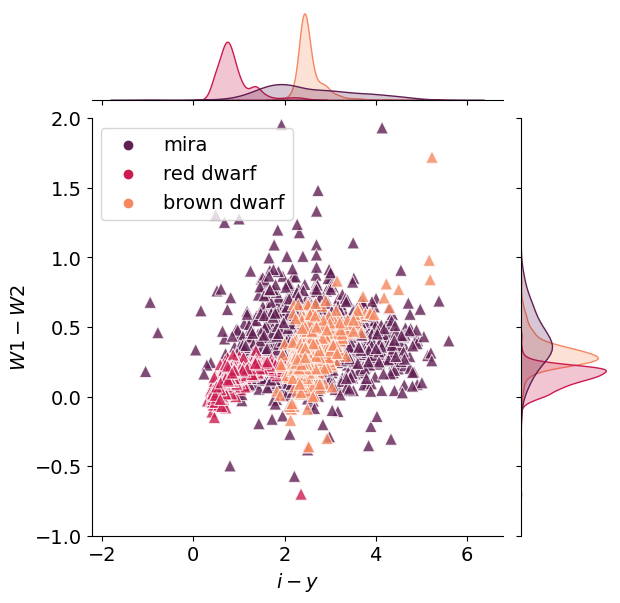}        
    \caption{Colour-colour diagrams with three types of objects: Mira variables (``mira''), red dwarfs and brown dwarfs. The points on the plots are overlapping, with the ``mira'' ones beneath everything else. On the marginal axes the separate density plots on each of three groups are shown.}
    \label{fig:fig1}
\end{figure}

Since brown dwarfs are usually too faint in the $i$, $z$, and $y$ bands of Pan-STARRS, we excluded objects with missing $i$, $z$, and $y$ magnitude values from consideration. Magnitudes in other bands are well-represented, and there are no missing values in our data for these bands. Although there is significant overlap between the loci of Mira variables and ultra-cool dwarfs, Mira variables often exhibit extremely red colors, likely due to extinction, as the reddest stars are located in the Galactic plane. To study this overlap in greater detail, we applied the following selection criterion to our dataset: $(i - W2) < 11^m$. We acknowledge that objects with $(i - W2) > 11^m$ belong to the Mira variable class in our data. For better representation, we randomly selected 1000 objects of each type-'brown dwarf', 'red dwarf' and 'Mira.' The colour-colour diagrams of the resulting dataset are presented in Fig.~\ref{fig:fig1} 

\section{Colour selection criteria}

We empirically devised the following selection criteria for distinguishing between the three classes:

i) For Mira variables: $(J - H) > 1.12$ \& $(i - W2) > 7.5$ or $(i - y) < 2.2$ \& $(J - H) > 0.8$ or $(i - W2) > 11$

ii) For red dwarfs: not i) and $(i - y) > 1.9$

iii) For brown dwarfs: not i) and not ii)

For differentiating between red dwarfs and brown dwarfs, we applied the rule established in \cite{2023A&C....4500744A} for Pan-STARRS $(i - y)$ colour.

\begin{figure}[ht]
    \centering
    \includegraphics[width=0.7\textwidth]{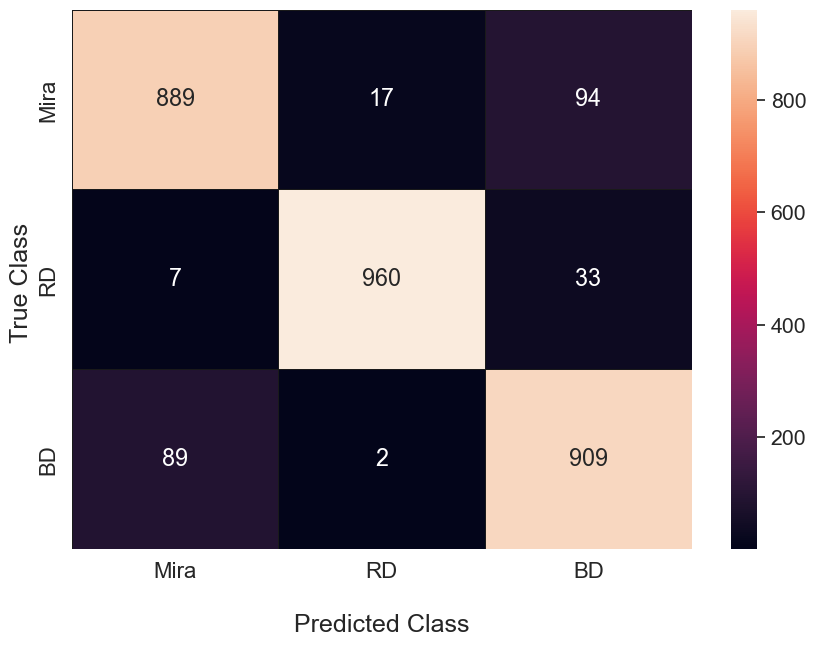}        
    \caption{Confusion matrix on the classification Mira variables, red dwarfs and brown dwarfs according to the rule established in this work. The true predictions are on the main diagonal.}
    \label{fig:fig2}
\end{figure}

We assigned the 'predicted' label according to the aforementioned rules and compared it to the 'true' labels. This comparison is shown in Fig.~\ref{fig:fig2} as a confusion matrix. Ideally, the result would have all three '1000' values on the main diagonal. While the classification of red dwarfs vs brown dwarfs and red dwarfs vs Mira variables works extremely well, the most significant challenges arise in differentiating between Mira variables and brown dwarfs. Nevertheless, even in this case, the efficiency of the classification remains quite impressive. We evaluate the effectiveness of the classification in terms of accuracy, which is defined as follows:

\begin{equation*}
    \textsc{Accuracy} = \frac{\textsc{True classifications}}{\textsc{Total}}
\end{equation*}

It is calculated as the sum of the values on the main diagonal divided by the sum of all values in the table, resulting in approximately $\approx 0.92$.

\begin{figure}[ht]
    \centering
    \includegraphics[width=0.5\textwidth]{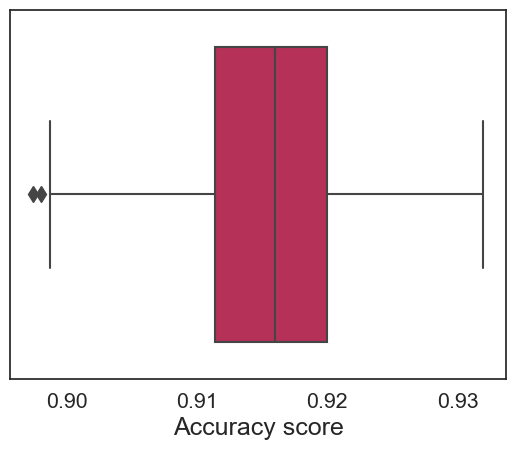}        
    \caption{The confidence interval of the accuracy calculated on 500 samples basis. }
    \label{fig:fig3}
\end{figure}

Since this exact value depends on the data, we evaluated it using a bootstrap technique. For the evaluation, we randomly selected a sample of 500 objects from each class (1500 in total) 500 times and calculated the accuracy in each case. The results are presented in a boxplot in Fig.~\ref{fig:fig3}. The colored box represents the range between the first quartile (Q1) and the third quartile (Q3), with the median value indicated by a black line. The error bars extend from the minimum to the maximum values, while outliers are highlighted as diamonds. As shown, the accuracy of the classification ranges from approximately 0.90 to 0.93.

\section{Conclusion}

This study presents a colour-based methodology for differentiating between Mira variables, red dwarfs, and brown dwarfs using data from WISE, 2MASS, and Pan-STARRS surveys. By analyzing the colour indices of these objects, we developed empirical selection criteria that effectively separate the three classes, achieving an overall classification accuracy of approximately 91\%-92\%. 

Our results demonstrate that while the classification of red dwarfs vs Mira variables is robust, distinguishing Mira variables from brown dwarfs presents a greater challenge due to their overlapping colour indices, particularly in cases where Mira variables exhibit moderate red colours. Despite these challenges, the developed criteria provide a reliable tool for photometric distinguishing of these classes. The results were confirmed by the bootstrap analysis, indicating that the classification accuracy remains consistently high across different subsamples. This work underscores ones again the importance of colour indices in stellar classification and offers a practical approach for separating these physically distinct yet observationally similar objects in large photometric surveys.



\bigskip

{\bf Acknowledgments:}
This work was funded by the Ministry of Science and Higher Education of the Russian
Federation, according to the research project 13.2251.21.0177 (075-15-2022-1228).
K.T. and A.L. are supported by the National Natural Science Foundation of China (NSFC) under grant No. 12261141689.
SJ and HPS acknowledge the financial support received from the BRICS grant DST/ICD/BRICS/Call-5/SAPTARISI/2023(G).



\bibliographystyle{spbasic}
\bibliography{mira}

\end{document}